# Coupling Model-Driven and Data-Driven Methods for Remote Sensing Image Restoration and Fusion

Huanfeng Shen, *Senior Member, IEEE*, Menghui Jiang, Jie Li, *Member, IEEE*, Chenxia Zhou, Qiangqiang Yuan, *Member, IEEE*, Liangpei Zhang, *Fellow, IEEE*

*Abstract*—**In the fields of image restoration and image fusion, model-driven methods and data-driven methods are the two representative frameworks. However, both approaches have their respective advantages and disadvantages. The model-driven methods consider the imaging mechanism, which is deterministic and theoretically reasonable; however, they cannot easily model complicated nonlinear problems. The data-driven methods have a stronger prior knowledge learning capability for huge data, especially for nonlinear statistical features; however, the interpretability of the networks is poor, and they are over-dependent on training data. In this paper, we systematically investigate the coupling of model-driven and data-driven methods, which has rarely been considered in the remote sensing image restoration and fusion communities. We are the first to summarize the coupling approaches into the following three categories: 1) data-driven and model-driven cascading methods; 2) variational models with embedded learning; and 3) model-constrained network learning methods. The typical existing and potential coupling methods for remote sensing image restoration and fusion are introduced with application examples. This paper also gives some new insights into the potential future directions, in terms of both methods and applications.**

## I. INTRODUCTION

REMOTE sensing data are carriers of spatial information and geographical knowledge, and the data quality directly affects the application, both intensively and extensively. However, the imaging process is influenced by many factors, such as the observation capability of the remote sensing satellite sensors, the land-cover type, the atmospheric conditions, and the lighting conditions, resulting in complex and diverse data quality problems [1].

The quality of remote sensing data is closely related to spatial, spectral, and temporal indicators. Specifically, the spatial resolution is the ability to discriminate the spatial detail information, which is the ground range corresponding to a pixel in the actual satellite observation image. The spectral resolution refers to the minimum wavelength interval that the sensor can resolve when receiving electromagnetic wave information radiated by a ground object. The temporal resolution is the revisit time of the sensor.

Due to the hardware limitations of the sensors, the energy that a sensor can receive is limited. A single remote sensing image needs to be a trade-off between the spatial, spectral, and temporal resolutions, leading to a low expression capability for the land surface. What is more, due to the interaction of sensor imaging, atmosphere transmission, and surface reflection, noise often degrades the spatial texture of remote sensing images. The two most typical noise types are the speckle noise in synthetic aperture radar (SAR) images and the hybrid noise in hyperspectral images (HSIs). Moreover, the thick cloud and thin cloud/haze appearing in poor atmospheric conditions can obscure the land, leading to missing spatial information and distorted spatial-spectral features.

Therefore, to overcome these degradation problems, researchers have proposed many image processing methods, including image denoising, cloud removal, and image fusion methods. The traditional remote sensing image processing methods are mostly based on filtering, regression, fitting, Fourier transform, and wavelet transform. However, these methods rarely consider the image degenerative process or the image priors existing in statistical and structural information. Therefore, variational methods have been proposed, which regard the image processing as an ill-posed inverse problem and construct the energy function according to the degradation models between the ideal image and the degraded observations. These methods are considered as model-based methods, in which the energy function is usually constructed based on Bayesian maximum *a posteriori* (MAP) estimation or sparse representation. These methods generally include two parts: a data fidelity model; and a prior model. The data fidelity model constrains the relationship between the ideal image and the degraded observations. The prior model employs the structure and statistical characteristics of the image itself to optimize the solution. Common image priors are the total variation (TV) prior [2], Laplacian prior [3], nonlocal prior [4], and low-rank prior [5]. Because of the rigorous theory, the accuracy of the variational model based methods is often higher than that of the traditional methods, but they cannot handle complicated nonlinear problems accurately.

Recently, deep learning has been applied to various remote sensing problems, due to its promising performance in describing the nonlinear relationships between different data. A variety of networks with good performances have been developed, such as ResNet [6], U-Net [7], encoder-decoder networks [8], DenseNet [9], and generative adversarial networks (GANs) [10]. However, although deep learning has powerful feature extraction and expression capabilities, it lacks a theoretical foundation and relies heavily on massive data. The model-driven methods (i.e., the variational model based methods) and the data-driven methods (i.e., the deep learning based methods) are complementary, to a large extent. On the one hand, the combination of the two methods can improve the interpretability of the network in deep learning and reduce the network's dependence on massive data; on the other hand, it can reduce the pressure of accurate modeling nonlinear problems of model-driven methods.

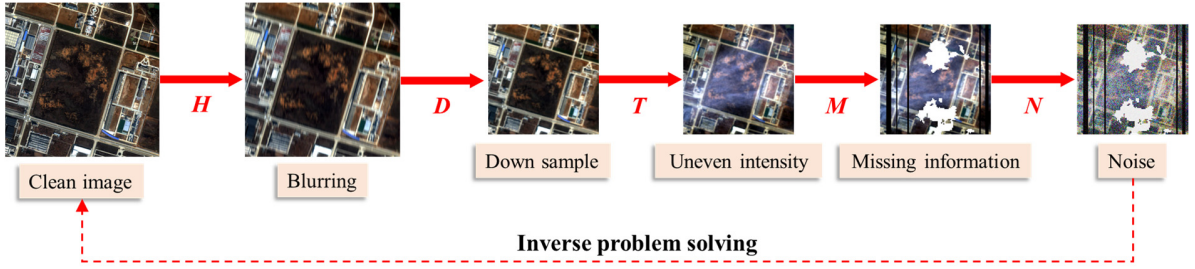

Fig. 1. The degradation processes of remote sensing images.

The coupling of model-driven and data-driven methods has been utilized in various computer vision tasks and has been found to be effective in image denoising [11], image super-resolution [12], image fusion [13], and image dehazing [14]. Along with the successful coupling of model-driven and data-driven methods in natural image processing [11]–[17], the coupling of these methods has also become a popular and promising trend in remote sensing. In this paper, we divide the coupling methods into three categories: 1) data-driven and model-driven cascading methods; 2) variational models with embedded learning; and 3) model-constrained network learning methods.

In the following, we first introduce the main idea behind each coupling approach, and we then introduce some specific applications of each coupling approach in the field of remote sensing image restoration and fusion. Although some attempts at the coupling of model-driven and data-driven methods have been made in remote sensing image restoration and fusion, there are still many gaps to be solved. We also discuss the future developments of data-driven and model-driven combinations in remote sensing.

The rest of this paper is organized as follows. Section II introduces the model-driven framework for remote sensing image restoration and fusion. In Section III, the ideas behind each coupling approach are described in detail. Section IV introduces some specific application examples. Finally, Section V and Section VI respectively provide the possible future developments and our conclusions.

## II. THE MODEL-DRIVEN FRAMEWORK

### A. Problem Definitions and Objective Functions in Remote Sensing

The degradation of remote sensing images often caused by noise, haze, cloud, and the lack of spatial-spectral-temporal resolution. The different degradation models in remote sensing can be uniformly expressed in the following framework:

$$Y = MTDHX + N \qquad (1)$$

where $Y$ is the observed degraded image, and $X$ is the ideal clean image. $T$, $D$, $H$, and $M$ represent the different degradation processes, as shown in Fig. 1. Among these processes, the downsampling matrix $D$ and the blur matrix $H$ are usually jointed to describe degradation of the imaging spatial scale. $T$ denotes an uneven variation of the intensity distribution, and it describes the portion of the electromagnetic radiation that reaches the sensor under the influence of complicated atmospheric scattering (due to thin cloud and haze) and the obstruction of the incident light. $M$ reflects the missing information, such as objects covered by thick cloud or dead pixels caused by sensor failure. $N$ denotes generalized noise, such as the classic Gaussian noise, Poisson noise, impulse noise and stripe noise.

To summarize $T$, $D$, $H$, and $M$ to be a matrix $A$, (1) can be simplified into:

$$Y = AX + N \qquad (2)$$

with the use of MAP estimation, the optimization problem can be formulated as:

$$X = \arg\min_X \|Y - AX\|_p^p + \lambda g(X) \qquad (3)$$

where the first term is the data fidelity term, and the second term is the regularization term. $g(X)$ in the second term is the prior operator, such as the TV operator, Laplacian operator, etc.

Based on sparse representation theory, the image restoration optimization problem can be rewritten as:

$$X = \arg\min_X \|Y - A\psi\alpha\|_2^2 + \lambda \|\alpha\|_0 \quad \text{with } X = \psi\alpha \qquad (4)$$

where the original clean image can be represented by the dictionary $\psi$ and the sparse coefficient $\alpha$ based on the spatial or spectral redundancy. The regularization term $g(\cdot)$ is also rewritten as some constraints associated to $\alpha$, as the sparsity prior of $l_0$ norm.

Furthermore, when improving the image quality with auxiliary data, such as image fusion with two observation images, the energy function model can be given as:

$$X = \arg\min_X \|Y - AX\|_p^p + \lambda g(X) + \gamma F_3(X, Z) \qquad (5)$$

where the third term is the data fidelity term between the ideal image $X$ and the other complementary observation $Z$.

However, for SAR intensity images with speckle noise, differing from (1) for optical data, the degradation model is usually described as:

$$Y = XN \qquad (6)$$

where $Y$ and $X$ are, respectively, the observed speckled image and the speckle-free image; and $N$ follows a Gamma law with one mean, which is signal-dependent multiplicative noise. Its density function is defined as:

$$p(N) = \frac{L^L}{\Gamma(L)} N^{L-1} e^{-LN} 1_{\{N \geq 0\}} \quad (7)$$

where $L$ is the number of looks. According to Bayesian MAP theory, SAR image despeckling can be regarded as an optimization problem [18], for which the model is as follows:

$$X = \arg \min_X \lambda (\log X + \frac{Y}{X}) + g(X) \quad (8)$$

In summary, the energy function of the image restoration problem can be expressed as:

$$X = \arg \min_X f(X, Y) + \lambda g(X) \quad (9)$$

where $f(\cdot)$ is the data fidelity term that keeps the consistency between the observed degraded images and the ideal clean image. The regularization term $g(\cdot)$ promotes solutions with an optimum performance, and parameter $\lambda$ controls the balance of consistency and performance. For the different tasks, $f(\cdot)$ and $g(\cdot)$ can be respectively used to describe different data relationships and prior constraints in (3), (5) and (8), which will be discussed in detail in Section IV.

### B. Basic Optimization Methods

For the solution of the above objective functions, two typical optimization strategies are often used. The simplest strategy is to solve the objective function directly by the gradient descent method, Newton's method, etc. Taking the case of gradient descent, (9) can be rewritten as:

$$X^{t+1} = X^t - \delta \frac{\partial}{\partial(X^t)}(f(X^t, Y) + \lambda g(X^t)) \quad (10)$$

where $t$ represents the iteration number, and parameter $\delta$ denotes the step size.

Another efficient strategy is the use of a variable splitting algorithm, such as half-quadratic splitting (HQS), the alternating direction method of multipliers (ADMM), or proximal gradient descent (PGD). In this strategy, an auxiliary variable $V$ is first introduced to split the data fidelity term and regularization term. Equation (9) can then be rewritten as:

$$X, V = \arg \min_{X,V} f(X, Y) + \lambda g(V), \text{ s.t. } AX + BV = C \quad (11)$$

The augmented Lagrangian of (11) is then:

$$X, V = \arg \min_{X,V} f(X,Y) + \lambda g(V) + \frac{\rho}{2} \|AX + BV - C + W\|_2^2 \quad (12)$$

where $W$ is the Lagrange multiplier, and $\rho$ is the penalty parameter.

Then, taking the ADMM as an example, to solve the above optimization problem, the iteration equations are:

$$X^{t+1} = \arg \min_X f(X, Y) + \frac{\rho}{2} \|AX + BV^t - C + W^t\|_2^2 \quad (13)$$

$$V^{t+1} = \arg \min_V \lambda g(V) + \frac{\rho}{2} \|AX^{t+1} + BV - C + W^t\|_2^2 \quad (14)$$

$$W^{t+1} = W^t + AX^{t+1} + BV^{t+1} - C \quad (15)$$

Finally, (13) and (14) can be solved in different ways, according to the specific tasks. When combined with deep learning, (14) can be implicitly replaced by a deep network prior with discriminative information for a specific problem.

## III. THE METHODS BASED ON THE COUPLING OF MODEL-DRIVEN AND DATA-DRIVEN

The coupling methods can be categorized into: 1) data-driven and model-driven cascading methods; 2) variational models with embedded learning; and 3) model-constrained network learning methods.

### A. Data-Driven and Model-Driven Cascading

The model-driven methods have difficulty in achieving accurate modeling, while the data-driven methods rely on massive data with representative features. Data-driven and model-driven cascading refers to the sequential use of data-driven and model-driven methods. This can be further categorized into two approaches (see also Fig. 2): i) first model-driven and then data-driven, i.e., first build a rough model to obtain an initialization result, and then use deep learning to generate a more accurate result; and ii) first data-driven and then model-driven, i.e., first use deep learning to mine the nonlinear features of the images, and then use the learned prior information to build an accurate model.

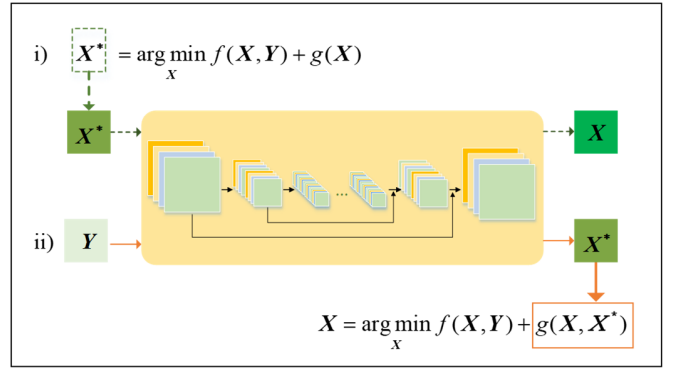

Fig. 2. The framework of data-driven and model-driven cascading.

As shown in Fig. 2, the dashed green arrow in the upper part indicates that the rough variational model, which utilizes the task-specific degradation domain knowledge, is first constructed to obtain a coarse reconstruction result, i.e., $X^*$. The coarse reconstruction result is then fed into the convolutional neural network (CNN) to allow the network to generate a more accurate reconstruction result, i.e., $X$.

The solid orange arrow in the lower part indicates that the CNN is first used to mine the deep prior information in the image, i.e., $X^*$. The prior is then used to construct an accurate energy function, which avoids complicated prior assumptions or incorrect linear assumptions. Equation (9) can then be rewritten as:

$$X = \arg \min_X f(X, Y) + \lambda g(X, X^*) \quad (16)$$

where $g(X, X^*)$ represents the relationship function between $X$ and $X^*$, such as the difference between the two or the difference between the gradients of the two. It should be noted that the learned prior image $X^*$ is not only used to construct the regularization term [13], [19], i.e., $g(X, X^*)$, but also can be used in the data fidelity term [20], i.e., $f(X, Y)$.

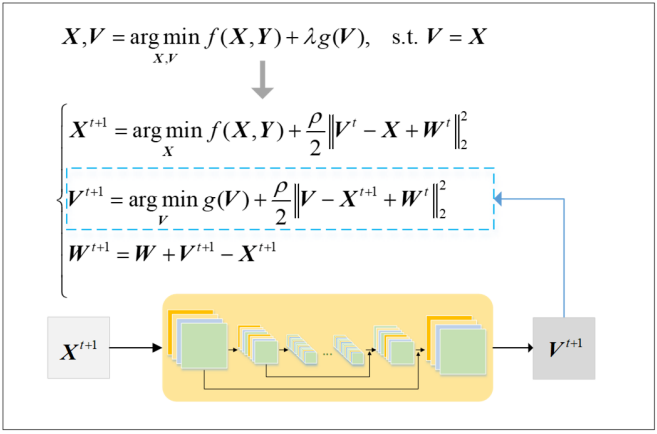

$$X, V = \arg\min_{X,V} f(X,Y) + \lambda g(V), \quad \text{s.t. } V = X$$

$$X^{t+1} = \arg\min_X f(X,Y) + \frac{\rho}{2}\left\|V^t - X + W^t\right\|_2^2$$

$$V^{t+1} = \arg\min_V g(V) + \frac{\rho}{2}\left\|V - X^{t+1} + W^t\right\|_2^2$$

$$W^{t+1} = W + V^{t+1} - X^{t+1}$$

Fig. 3. The framework of variational models with embedded learning.

### B. Variational Models with Embedded Learning

Both the handcrafted priors in the traditional variational models and the deep priors in data-driven and model-driven cascading approaches explicitly define the regularization term. However, the degradation types of remote sensing images are complex and diverse, and an explicit prior cannot handle the various latent degradation types. Thus, the plug-and-play prior strategy [21], which is known for its flexible and effective handling of various inverse problems, is used. The main idea of this approach is to unfold the energy function into sub-problems by the use of a variable splitting algorithm, and then to embed the pre-trained CNN to solve the prior term related sub-problem.

As shown in Fig. 3, the auxiliary variables $V$ introduced by the ADMM algorithm decouple the data fidelity term and the regularization term of the model into individual sub-problems. The variable $X$ of the data fidelity term related sub-problem is solved using a conventional solution, such as gradient descent or the least-squares algorithm. The variable $V$ of the regularization term related sub-problem is solved through the plug-and-play pre-trained network.

Most of the existing plug-and-play prior based image restoration methods treat the CNN Gaussian denoiser as the prior [11,22], and some treat the CNN super-resolver as the prior [12].

### C. Model-Constrained Network Learning

The previous two approaches are aimed at capturing a more accurate prior by the data-driven method, and then the pre-trained prior is used to construct the energy function or solve the prior-related sub-problem. In these two approaches, the network training and model solving are separated. In contrast, the model-constrained network learning approach integrates the model-driven and data-driven methods into an end-to-end network, simultaneously performing network parameter optimization and model solving. According to the different alternatives, the model-constrained network learning approach can be further divided into two subcategories: 1) model-constrained network structure; and 2) model-constrained loss function. The former uses the model to make the network structure interpretable, and the latter uses the model to constrain the optimization space of the network parameters.

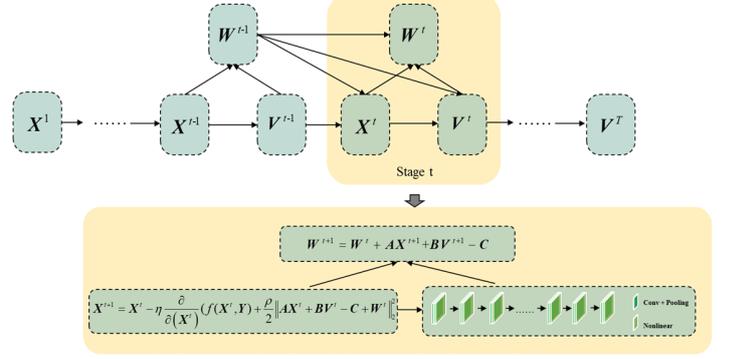

Fig. 4. The framework of model-constrained network structure.

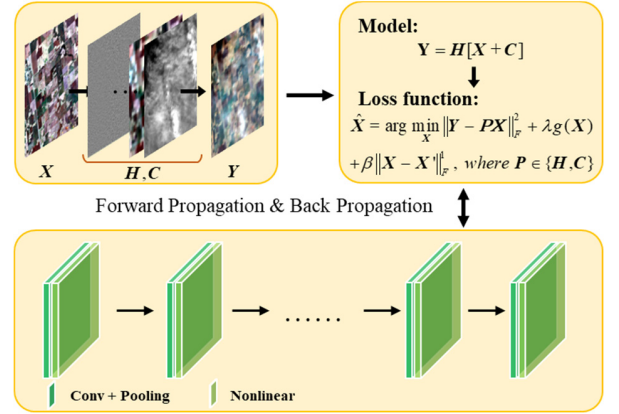

Fig. 5. The framework of model-constrained loss function.

#### 1) The Model-Constrained Network Structure

As known, variable splitting and alternate iterative optimization algorithms are typically used for optimizing the solutions of the model-driven methods. However, the processing of the optimization has to manually adjust some tedious parameters, and only obtain the shallow feature priors. Hence, in order to avoid these problems, this approach is to introduce a recursive network structure to realize autimatic parameters updating and deep feature mining. Each iteration in the orignal model-driven methods will be unfold into one sub-network in recursion. The multi-cascaded sub-network form into a recursive network structure, in which sub-networks can share or do not share the same parameters [15,16,23-26].

Similar to the variational models with embedded learning, a variable splitting algorithm, taking the ADMM algorithm as an example, first decouples the energy function (9) into individual sub-problems, such as (13)–(15). The network is then designed according to the iterative update processes of the sub-problems, as shown in Fig. 4. A deep convolutional structure (DCS) is considered to solve at least one sub-problem. Fig. 4 takes the solution of a sub-problem as an example [15], where the whole network contains T stages, corresponding to T iterative processes. At each stage, the sub-variables are updated in turn, the DCS solves the regularization term, and the data fidelity term is solved by the traditional method, which is given by:

$$
\begin{cases}
\boldsymbol{X}^{t+1} = \boldsymbol{X}^t - \eta \dfrac{\partial}{\partial(\boldsymbol{X}^t)}(f(\boldsymbol{X}^t,\boldsymbol{Y}) + \dfrac{\rho}{2}\left\|\boldsymbol{A}\boldsymbol{X}^t + \boldsymbol{B}\boldsymbol{V}^t - \boldsymbol{C} + \boldsymbol{W}^t\right\|_2^2) \\
\boldsymbol{V}^{t+1} = DCS(\boldsymbol{X}^{t+1},\boldsymbol{W}^t) \\
\boldsymbol{W}^{t+1} = \boldsymbol{W}^t + \boldsymbol{A}\boldsymbol{X}^{t+1} + \boldsymbol{B}\boldsymbol{V}^{t+1} - \boldsymbol{C}
\end{cases}
\tag{17}
$$

All parameters are jointly optimized by an end-to-end training scheme. Natually, other optimization methods [16, 23-26] can also be applied to solve this category. For example, ISTA-NET [16] and the dual-path deep neural network (DPDNN) [23] were inspired by the iterative shrinkage-thresholding algorithm (ISTA) and HQS, respectively. Most of these methods employed a DCS for the optimization of the regularization term while using the gradient descent method or Newton's method for the optimization of the data fidelity term, which was embedded in the network with an equation. However, in [24], Kobler *et al.* proposed a variational network that employed the PGD method, and used a rectified linear unit (ReLU) instead of proximal mapping.

### 2) Model-Constrained Loss Function

The commonly used loss functions in image restoration tasks, such as the L1 and L2 norms, focus on minimizing the difference between the network output and label data. However, they do not consider the physical model behind the degradations, and thus cannot be categorized as model-driven. In this paper, the model-constrained loss function is established by integrating some degenerate relationships between observed image and estimated image, and then can be regarded as a network parameter optimization problem with degradation model constraint, as shown in (18). The framework of this approach is shown in Fig. 5.

$$
\begin{aligned}
\hat{\boldsymbol{X}} &= \arg\min_{\boldsymbol{X}} f(\boldsymbol{X},\boldsymbol{Y}) + \lambda g(\boldsymbol{X}) + \beta R(\boldsymbol{X},\boldsymbol{X}') \\
&= \arg\min_{\boldsymbol{X}} \left\|\boldsymbol{Y} - \boldsymbol{P}\boldsymbol{X}\right\|_F^2 + \lambda g(\boldsymbol{X}) + \beta \left\|\boldsymbol{X} - \boldsymbol{X}'\right\|_F^1
\end{aligned}
\tag{18}
$$

where $\left\|\cdot\right\|_F$ denotes the Frobenius norm. $\lambda$ and $\beta$ are predefined balance scalars. The energy function $f(\boldsymbol{X},\boldsymbol{Y})$ can be defined according to the different application tasks, indicating the multiplicative and additive degenerations in the spectral and spatial domains. For example, the noise that obeys a Gamma distribution in SAR imagery, the downscaling in the super-resolution problem, the blur kernel in deblurring, the thin cloud/haze in optical images, and the noise that follows a Bayes model. $\boldsymbol{X}$ is the output of the network, and $\boldsymbol{X}'$ is the label data of $\boldsymbol{X}$. $\boldsymbol{Y}$ indicates the observed image with one or multiple degradation factors. $g(\boldsymbol{X})$ is the handcrafted prior term, such as TV prior, nonlocal prior, etc. Traditionally, $R(\boldsymbol{X},\boldsymbol{X}')$ is the mean-squared error (MSE), which acts as a loss function to constrain the relationship between the label data and the estimated solution. Designing loss function of the network according to $\boldsymbol{P}$ and the handcrafted prior $g(\boldsymbol{X})$ can effectively reduce the optimization space of the network parameters and improve the accuracy of the network.

## IV. APPLICATION EXAMPLES

### A. Speckle Noise Reduction in SAR Images

SAR systems are active remote sensing systems, which play an important role in Earth surface monitoring, irrespective of weather conditions. However, in SAR images, the speckle noise has the characteristics of multiplicative Gamma noise, and inherently affects the image quality, making the imagery hard to classify, interpret, or segment, due to the coherent interference of radar waves reflected from the many basic scatterers. Therefore, the reduction of speckle noise in SAR imagery is important for many applications. Generally speaking, the various despeckling approaches can be categorized as model-driven or data-driven methods. The coupling model-driven and data-driven methods are few in number. In the following, we introduce some examples for the two categories [26-29], i.e., the variational models with embedded learning and the model-constrained network learning approaches.

### 1) Despeckling With the Variational Models With Embedded Learning Approach

In the class of the variational models with embedded learning, there are mainly three combination modes for single-polarization SAR images despeckling, i.e., in the original intensity domain, in the log intensity domain, and in the original complex domain. In 2017, Deledalle et al. [27] proposed a plug-and-play SAR image reconstruction framework in the log intensity domain, in which the multiplicative noise is transformed into additive noise by log-transform. And then, in 2019, Alver et al. [28] proposed a different framework in the original complex domain, in which the authors assume a Fourier transform-based forward model for constructing the additive noise observation mode.

There is no relevant paper in the original intensity domain based on the original multiplicative observational model now. Therefore, for the first time, we proposed a novel method based on the original multiplicative degradation model of (7), called SAR-PNP in this reviewer paper. Here, we use the AA variational model proposed by Aubert and Aujol [18], which is presented in (8). Then, taking the variable splitting technique and HQS algorithm, the iteration equations are as follows:

$$
\boldsymbol{X}^{t+1} = \arg\min_{\boldsymbol{X}} \lambda\,(\log \boldsymbol{X} + \frac{\boldsymbol{Y}}{\boldsymbol{X}}) + \frac{\rho}{2}\left\|\boldsymbol{X} - \boldsymbol{V}^t\right\|_2^2
\tag{19}
$$

$$
\boldsymbol{V}^{t+1} = \arg\min_{\boldsymbol{V}} g(\boldsymbol{X}) + \frac{\rho}{2}\left\|\boldsymbol{V} - \boldsymbol{X}^{t+1}\right\|_2^2
\tag{20}
$$

where the sub-problem $\boldsymbol{V}$ is solved by the deep CNN network $\boldsymbol{V}^{t+1} = DCNN(\boldsymbol{X}^t)$, which is a pre-trained network, and the sub-problem $\boldsymbol{X}$ is solved by the gradient descent method.

### 2) Despeckling With the Model-Constrained Network Structure Approach

As the example of the model-constrained network structure approach for SAR despeckling, Shen *et al.* [26] proposed a recursive deep convolutional neural network (DCNN) prior model (SAR-RDCP), which jointly optimizes the data fidelity term and the deep CNN based regularization term. In this method, Shen *et al.* apply the same framework presented in (19) and (20). The difference is the strategy of the network learning. Specifically, the two sub-problems $X$ and $V$ are optimized and updated jointly in a recursive network structure instead of only updating the sub-problem $V$ in network.

To validate the effectiveness of the variational models with embedded learning approach and the model-constrained network learning approach, the real-data experiments were undertaken. Therefore, the deep learning based methods, such as the pure network method called SAR-DCNN [26], which is the same as the structure of DCNN in SAR-PNP and SAR-RDCP, are selected for comparison with SAR-PNP and SAR-RDCP. Here, the UC Merced land-use dataset [30] is used as training set for the deep learning based methods. A Sentinel-1 single look complex (SLC) VH format image of the city of Wuhan in China was selected for the experiment, as shown in Fig. 6. The image was cropped to $500 \times 500$ pixels for the experiment. From the results, it can be seen that the detail preservation and speckle reduction of the SAR-RDCP method and the SAR-PNP method are superior than the SAR-DCNN method. For the SAR-PNP and SAR-RDCP methods, from the two enlarged images of Figs. 6(c) and 6(d), the place selected in the red box indicates that the SAR-PNP method retains the details of the water body area better, but from the place selected in the blue box, the SAR-RDCP method has better retention of strong scattering points.

From the results described above, the significant improvement of the image quality obtained using the SAR-RDCP and SAR-PNP methods shows that the use of the model-constrained network learning structure with optimizing guidance and the use of a pre-trained individual network may have greater potential than the pure network. In addition, the two methods show different advantages in different scenes. Thus, for different scenes, it is also meaningful to choose appropriate coupling methods to obtain better solution.

Recently, Molini *et al.* [29] applied a Bayesian framework relying on blind-spot CNNs to the self-supervised SAR image despeckling task. The main idea of this work is to minimize the negative logarithm likelihood distribution in the training phase. Motivated by the method, it can be observed that a model-constrained loss function with appropriate likelihood distribution can also makes a breakthrough that takes into account the characteristics of the speckle in the network optimization.

Overall, the coupling methods are being more widely used in SAR image despeckling, and obtain satisfactory performances. Up to now, these methods are mainly for single-polarization SAR regardless of full-polarization SAR. The extreme lack of real clean full-polarization SAR imagery make us put more attention to the unsupervised strategies, in which a variational model better help to constrain both the network structure and loss function.

### B. HSI Denoising and Reconstruction

Hyperspectral imaging is a technique used to acquire the radiation characteristics of the observed objects with a fine spectral resolution. On account of the rich spectral information, HSIs are utilized in many applications. However, the increase in the spectral channels of the sensors generates spectra with low signal-to-noise ratios. Due to the observation conditions and sensors, HSIs are always degraded by multiple types of noise, such as Gaussian noise, stripe noise, and impulse noise. On the other hand, HSIs are also compressed to avoid much pressures to data storage, transmission and processing of airborne or spaceborne remote sensing imaging system. Therefore, HSI denoising and reconstruction are the common tasks of recovering a clean HSI from its noisy and compressed versions.

As known, the data-driven methods have shown better performances than the conventional model optimization based methods due to their powerful representation capability; however, these methods lack flexibility. Hence, few methods combining the data-driven and model-driven approaches have been attempted for HSI denoising and reconstruction, and have shown advantages in exploiting the large training datasets and introducing the explicable structure of the prior-regularized optimization. These coupling methods are mainly based on the variational model with embedded learning approach and model-constrained network learning.

### 1) HSI Denoising With the Variational Models With Embedded Learning

For the variational models with embedded learning, plug-and-play frameworks are the main approach used in HSI denoising. These frameworks provide us with the possibility of integrating the capabilities of multiple priors, including deep learning priors, in one restoration model. For example, Zeng *et al.* [31] embedded a low-rank prior and a deep learning prior into a plug-and-play framework. This method uses a sub-model of Tucker decomposition based low-rank tensor approximation to remove the sparse noise and part of the Gaussian noise, and leaves the residual by the dilated deep residual network. Otherwise, a spectral mixing model can also be integrated into the framework, and a deep learning prior can be used to excavate the implicit features of the abundance matrix [32]. In general, this type of approach can be generalized as:

$$\arg\min_{X} \left\| Y - X + S + N \right\|_F^2 + \tau g(Z) + \lambda \left\| S \right\|_1 + \beta \left\| N \right\|_F^2 \quad (21)$$
$$\text{s.t. } X = Z$$

where $\left\| Y - X + S + N \right\|_F^2$ is equivalent to $f(\cdot)$ in (9), and $g(\cdot)$ denotes the implicit CNN-based function. $N$ and $S$ respectively denote the Gaussian noise and the sparse noise, which consists of impulse noise, stripes, deadlines, etc. As in [32], HSIs can also be transformed into signal subspaces for mining deeper characteristic relationships, e.g., sparse representation in the spectral dimension, with $X = EA$, where $E$ and $A$ are, respectively, the endmember matrix and abundance matrix. $\lambda$ and $\beta$ are positive regularization parameters.

*2) HSI Reconstruction With the Model-Constrained Network Learning Approach*

The model-constrained network learning approach is currently used in HSI reconstruction. Wang *et al.* [33] first introduced iterative optimization into a deep convolutional network. This method focuses on exploiting the spatial and spectral correlation, which can lead to a superior performance. Firstly, the optimization problem, which guarantees the relationship between the desired HSI and the original HSI, is unfolded into an iteration-based optimization problem, as shown in (17), by the use of the conjugate gradient algorithm. Secondly, the structural insight of the iterative processing is integrated into the network and forms a data-driven prior, which can also be called an optimization-inspired network. The data-driven prior can regularize the optimization problem to exploit the spatial and spectral correlation, thus removing the influence of the noises and avoiding the heavy computational load of the traditional iterative optimization methods. The optimization inspired algorithm can be given as:

$$\hat{X} = \arg\min_{X} \| Y - \Phi X \|_F^2 + \tau g(H), \text{ s.t. } H = X$$

$$\Rightarrow \begin{cases} X^{(k+1)} = X^{(k)} - \varepsilon \left[ \Phi^T \left( \Phi X^{(k)} - Y \right) + \eta \left( X^{(k)} - H^{(k)} \right) \right] \\ H^{(k)} = g(X) \end{cases} \quad (22)$$

where the auxiliary variable $H$ is introduced to convert the optimization problem into a sub-problem about $H$ related to the hyperspectral image prior $g(\cdot)$. Compared to reconstruction problem, HSI denoising is more basic degradation problem. Naturally, when $\Phi$ is identity matrix, equation (22) can be converted to deal with noises in HSIs.

*3) HSI Denoising With the Model-Constrained Loss Function Approach*

In the model-constrained loss function approach, the image prior [34] can be found directly in the space of the network's parameters, through the optimization process with loss function considering degeneration relationship between input and output images, which can be called degeneration loss function in this review. This function called deep hyperspectral prior (DHSP) can describe the fidelity between the estimated and corrupted images in denoising, inpainting, super-resolution, and can be written as:

$$\hat{\theta} = \arg\min_{X} E\left( f_\theta(Y), Y \right), \text{ s.t. } X = f_\theta(Y) \quad (23)$$

Similar to (23), to utilize the convolutional autoencoder [35] in the training processing, an HSI can also be converted into a nonlinear representation $\alpha$ by using the encoder function. The HSI can then be reconstructed from $\alpha$ using the decoder function, which therefore acts as an HSI prior for the compressive reconstruction. Considering the non-i.i.d. noise of HSIs, a denoising framework called the deep spatio-spectral Bayesian posterior (DSSBP) method [36] for non-i.i.d. noise was designed based on the deep spatio-spectral Bayesian posterior in HSIs, which models the HSI non-i.i.d. noise across different bands [37], and can obtain a good performance.

To analyze the effectiveness of the variational model with embedded learning approach and model-constrained network learning approach, some representative methods are chosen in the following experiments. These methods are respectively the deep learning model FFDNet[1] [38], the deep plug-and-play model with the combination of FFDNet and LRTA (DPLRTA[1]) [31], and DHSP[2] [34] that is the model-constrained loss function based method. The Hyperspectral Digital Imagery Collection Experiment (HYDICE) Urban dataset with the cropped size of $256 \times 256 \times 191$ was adopted in the simulated HSI data denoising experiments to interpret the performance of different approaches. As the DHSP based method is mainly proposed to remove the Gaussian noise, $\sigma_b = 25$ is only added to each band, and these images before and after restoration are displayed as false color with $57th$, $27th$ and $17th$ bands in Fig. 7. From the results, it is apparent that all are able to remove the Gaussian noise, but DHSP can better maintain the spectral characteristic than other while the image restored by the DPLRTA shows the good structure and texture information. With the mean peak-signal-to-noise ratio (PSNR), the mean structural similarity (SSIM) index, the mean spectral angle (MSA) mapper served as evaluation indices, the quantitative assessment also indicates the DPLRTA obtain the best PSNR and SSIM, and DHSP gives the best MSA. In addition, authors in [31] have proven that DPLRTA can remove the large-scale sparse noise compared with FFDNet. Certainly, here as an unsupervised method, DHSP can obtain an image prior within a CNN itself and flexibly adapt for situation without an amount of training data available. Certainly, to overcome the hybrid noise as DPLRTA, the loss function should be adjusted to describe the properties of different noises, rather than the only mean square error function for Gaussian noise.

Overall, compared with the variational models with embedded learning, the model-constrained network learning approach can adaptively manage the hyperparameter learning in the network training process, instead of manually adjusting the regularization parameters, and can obtain better convergence results for nonlinear problems. Conversely, the variational models with embedded learning can be flexibly adjusted for different types of noise by adding different prior models. However, the accuracy and training speed of the HSI denoising using the coupled model-driven and data-driven methods are still needed to be considered.

*C. Remote Sensing Image Fusion*

Due to the limitations of the hardware, remote sensing images are trade-offs between the spatial resolution and spectral resolution. Remote sensing image fusion refers to the fusion of multiple images with complementary information to obtain higher-resolution remote sensing images. Here, we refer to spatial-spectral fusion, which aims at obtaining a fused image with both high spatial and spectral resolutions by fusing high spatial resolution/low spectral resolution images and low spatial resolution/high spectral resolution images. This includes panchromatic (PAN)/multispectral image (MSI) fusion, PAN/

---

[1] https://github.com/NavyZeng/DPLRTA

[2] https://github.com/acecreamu/deep-hs-prior

HSI fusion, and MSI/HSI fusion. PAN/HSI fusion can be regarded as a special case of MSI/HSI fusion.

For remote sensing image fusion, the energy function can be generally represented as (5), where $\boldsymbol{X} \in \mathbb{R}^{WH \times S}$ is the ideal high spatial-spectral resolution image. W, H, and S are the width, height, and band number of the ideal image. $\boldsymbol{Y} \in \mathbb{R}^{wh \times S}$ denotes

TABLE I
APPLICATIONS OF THE COUPLED MODEL-DRIVEN AND DATA-DRIVEN METHODS

| Reference | Category | Application |
|---|---|---|
| Shen *et al.* [26] | model-constrained network structure | SAR despeckling |
| Deledalle *et al.*[27] | variational models with embedded learning | SAR despeckling |
| Alver *et al.* [28] | variational models with embedded learning | SAR despeckling |
| Zeng *et al.* [31] | variational models with embedded learning | HSI denoising |
| Lin *et al.* [32] | variational models with embedded learning | HSI denoising |
| Choi *et al.* [35] | variational models with embedded learning | HSI denoising |
| Wang *et al.* [33] | model-constrained network structure | HSI reconstruction |
| Sidorov *et al.* [34] | model-constrained loss function | HSI denoising |
| Zhang *et al.* [36] | model-constrained loss function | HSI denoising |
| R. Dian *et al.* [13] | data-driven and model-driven cascading | MSI/HSI fusion |
| W. Xie *et al.* [19] | data-driven and model-driven cascading | PAN/HSI fusion |
| H. Shen *et al.* [20] | data-driven and model-driven cascading | PAN/MSI fusion |
| R. Dian *et al.* [22] | variational models with embedded learning | MSI/HSI fusion |
| Q. Xie *et al.* [47] | model-constrained network structure | MSI/HSI fusion |
| L. Zhang *et al.* [48] | model-constrained loss function | MSI/HSI fusion |

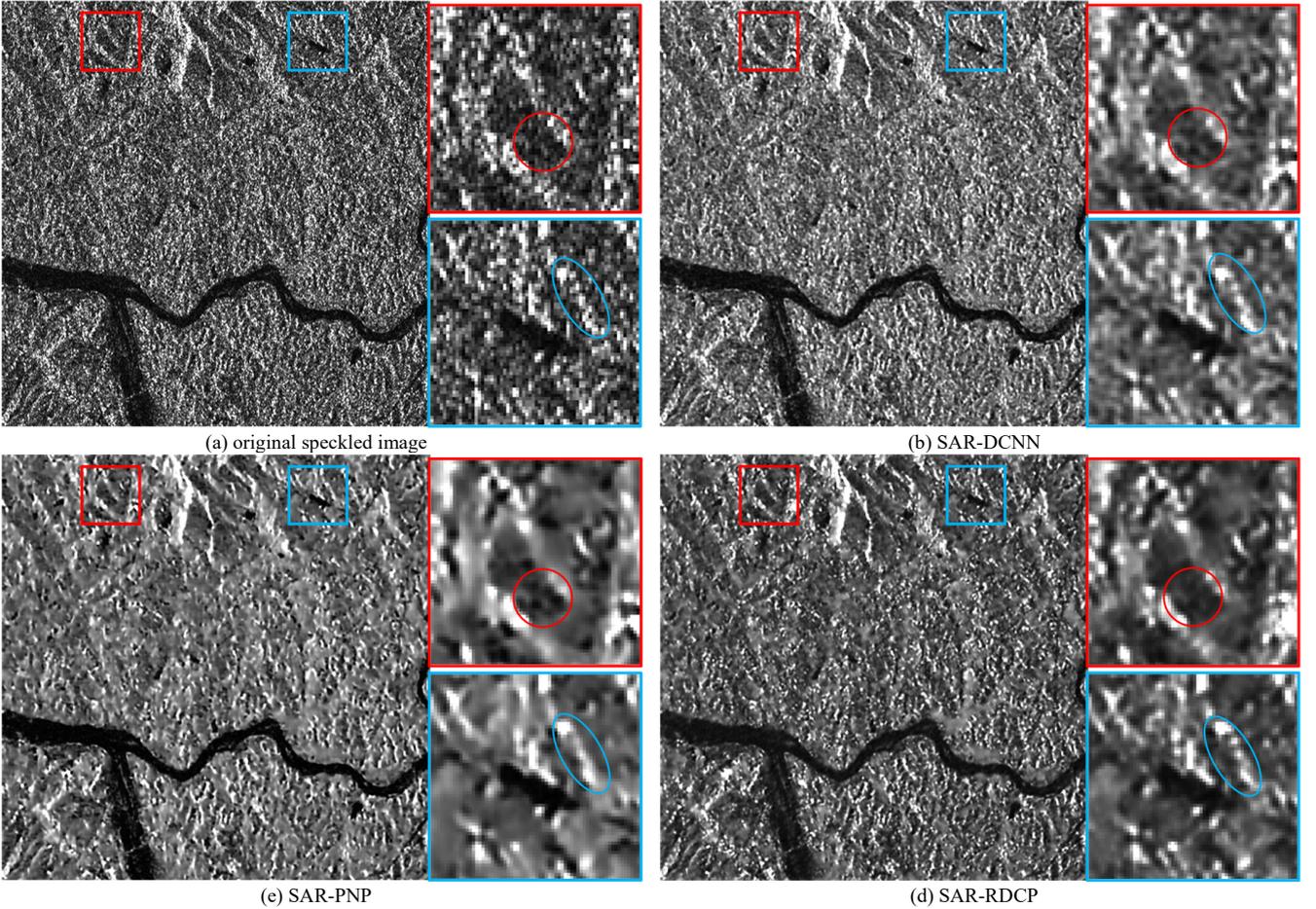

(a) original speckled image

(b) SAR-DCNN

(e) SAR-PNP

(d) SAR-RDCP

Fig. 6. Despeckling results for the Sentinel-1 VH SLC image of the city of Wuhan. (a) Original speckled image (b) SAR-DCNN. (c) SAR-PNP. (d) SAR-RDCP.

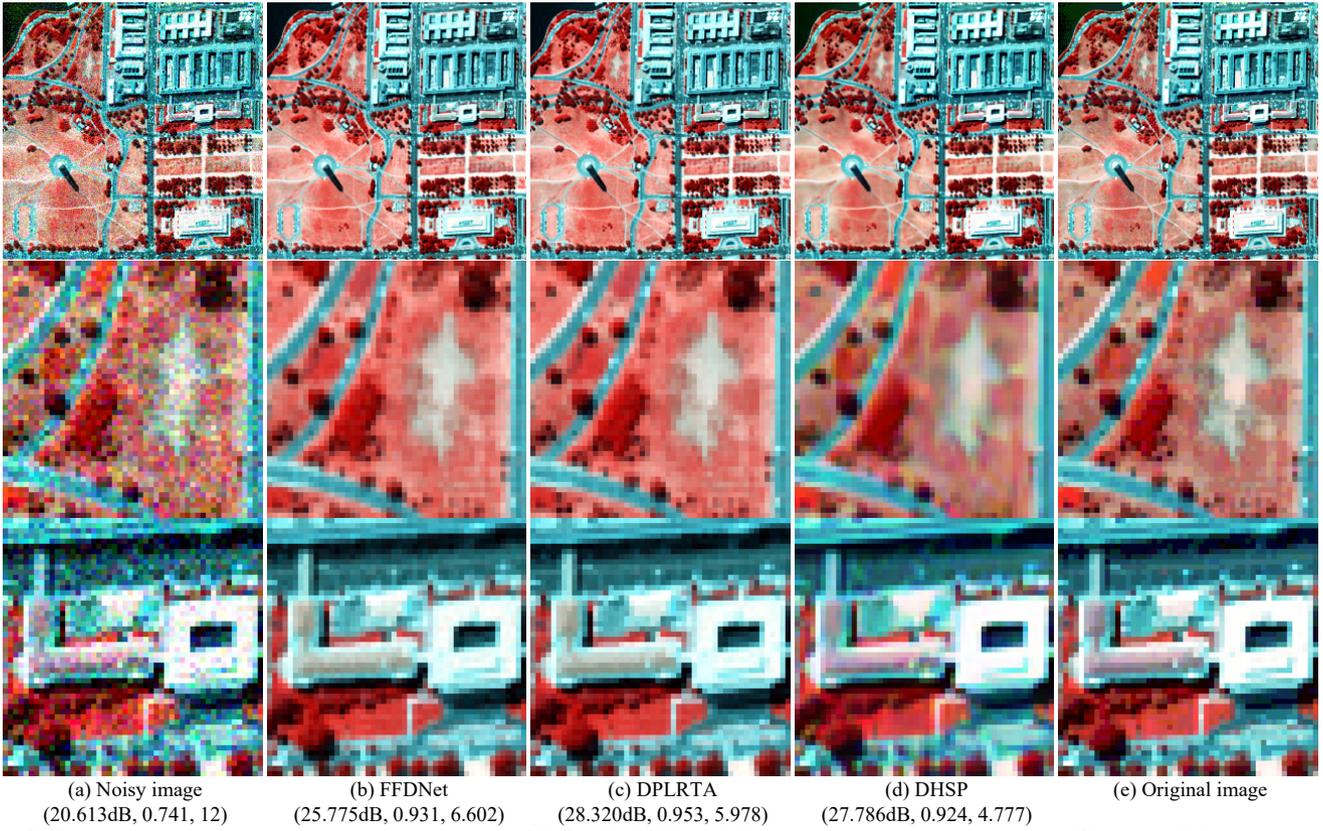

(a) Noisy image (20.613dB, 0.741, 12)    (b) FFDNet (25.775dB, 0.931, 6.602)    (c) DPLRTA (28.320dB, 0.953, 5.978)    (d) DHSP (27.786dB, 0.924, 4.777)    (e) Original image

Fig. 7. The visual comparision and quantitative evaluation with PSNR (dB), SSIM and MSA values of the denoised results in the Washington DC mall dataset: (a) Noisy image (57, 27, 17), (b) FFDNet, (c) DPLRTA, (d) DHSP, (e) Original image.

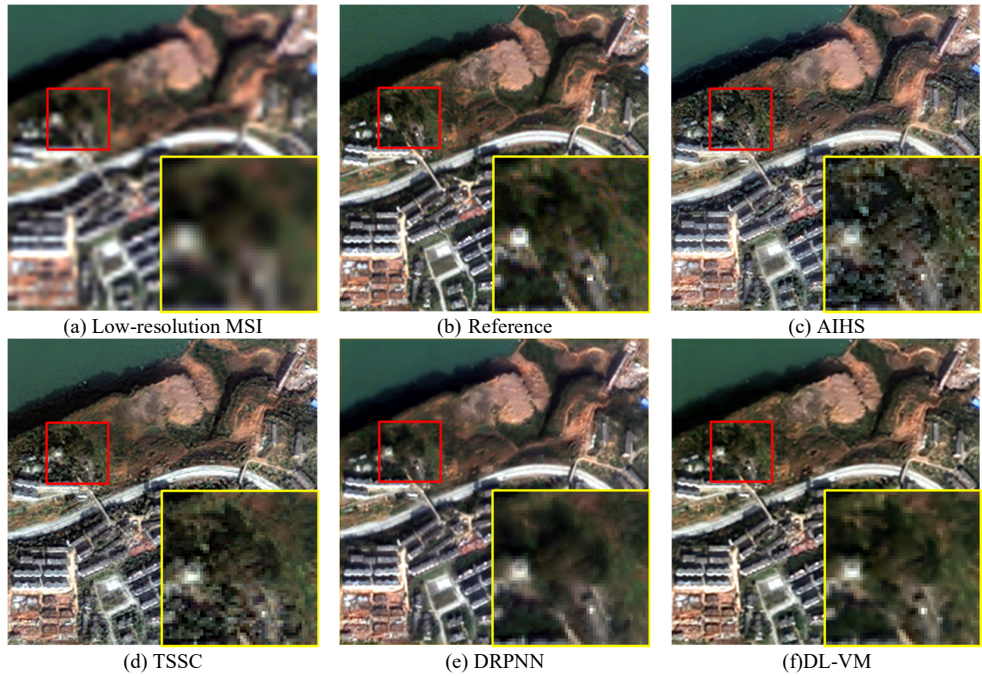

(a) Low-resolution MSI    (b) Reference    (c) AIHS

(d) TSSC    (e) DRPNN    (f)DL-VM

Fig. 8. Fusion results for the QuickBird MSI and PAN images: (a) low-resolution MSI image, (b) reference image, (c) MTF-GLP-HPM, (d) TSSC, (e) DRPNN, and (f) DL-VM

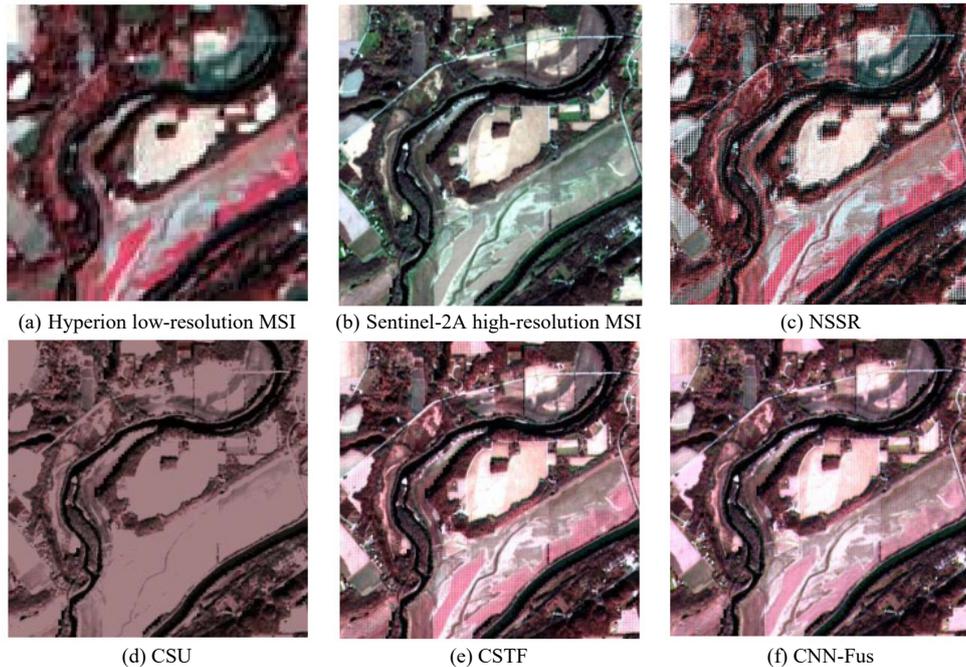

(a) Hyperion low-resolution MSI     (b) Sentinel-2A high-resolution MSI     (c) NSSR

(d) CSU     (e) CSTF     (f) CNN-Fus

Fig. 9 [22]. Fusion results for the Hyperion MSI and Sentinel-2A MSI images: (a) Hyperion low-resolution MSI, (b) Sentinel-2A high-resolution MSI, (c) NSSR, (d) CSU, (e) CSTF, and (f) CNN-Fus.

the observed low spatial resolution/high spectral resolution image. W/w is the spatial resolution ratio of the corresponding $X$ and $Y$ . $Z \in \mathbb{R}^{\text{WH} \times \text{s}}$ is the observed high spatial resolution/low spectral resolution image, with $\text{s} \ll \text{S}$ . The first term is the spectral fidelity term, where $H \in \mathbb{R}^{\text{wh} \times \text{HW}}$ is the downsampling and blurring matrix. The third term is the spatial fidelity term.

In remote sensing fusion, scholars have proposed methods belonging to each coupling class. In the following, we introduce representative fusion methods for each coupling class in detail.

*1) Fusion With the Data-Driven and Model-Driven Cascading Approach*

As an example of the data-driven and model-driven cascading approach, Shen *et al.* [20] proposed the DL-VM method, which utilizes the gradient prior obtained from a CNN to construct the spatial fidelity term for PAN/MSI fusion, and it is the first method combining data-driven and model-driven methods for remote sensing image fusion. The energy function of [20] is written as:

$$X = \arg\min_{X} \frac{1}{2} \|Y - HX\|_2^2 + \frac{\gamma}{2} \sum_{j=1}^{2} \|\nabla_j X - G_j\|_2^2 + \frac{\lambda}{2} \|QX\|_2^2 \ (24)$$

The second term is the spatial fidelity term, where $G_j \in \mathbb{R}^{\text{WH} \times \text{S}}$ with $j = 1, 2$ denotes the gradient images of the high-resolution MSI in the horizontal and vertical directions learned through the network, corresponding to $X^*$ in Fig. 2. $\nabla_j \in \mathbb{R}^{\text{WH} \times \text{WH}}$ with $j = 1, 2$ means the global first-order finite difference matrix in the horizontal and vertical directions, respectively. Constructing the spatial fidelity term with the learned gradient priors $G_j$ avoids the inaccurate linear assumption of the relationship between the PAN and high-resolution MSI. $Q \in \mathbb{R}^{\text{MN} \times \text{MN}}$ in the third term indicates the Laplacian matrix, which is a common Laplacian prior term.

In this class, in addition to pansharpening, Dian *et al.* [13] proposed DHSIS for MSI/HSI fusion, Xie *et al.* [19] proposed HPDP for PAN/HSI fusion. The data-driven and model-driven cascading strategy uses the prior learned from a pre-trained network to construct the energy function, which avoids the linear assumption in the spatial fidelity term or the regularization term of the classic model-driven method. However, some limitations exist in this strategy: 1) it requires retraining of the network when fusing data of different sensors; and 2) it takes time to iteratively solve the energy function.

To show the effectiveness of the data-driven and model-driven cascading approach in PAN/MSI fusion, a low-resolution QuickBird MSI with the size of $248 \times 248 \times 4$ and a QuickBird PAN image with the size of $992 \times 992 \times 1$, with resolutions of 2.44 m and 0.61 m, respectively, were adopted in the simulated experiment according to the Wald's protocol. The traditional component substitution-based method of AIHS [39], the model-based TSSC method [40], and the deep learning based method of DRPNN [41] were applied to compare with the cascading method DL-VM [20]. Fig. 8 shows the R-G-B band combinations of the various fusion methods, where the lower right corner is a magnified display of the image inside the red rectangle. By comparing the results, it can be observed that, for AIHS and TSSC, sharpened spatial features are achieved, but with severe spectral distortion, as can be seen in the vegetation area in Figs. 8(c)–(d). DRPNN shows a good performance in spectral fidelity, but performs poorly in spatial texture information enhancement, as can be seen in the zoomed area in Fig. 8(e). The fusion result obtained by DL-VM is the closest to the reference image, both in the fusion of the spatial details and in the preservation of the spectral fidelity.

### 2) Fusion With the Variational Models With Embedded Learning Approach

In the variational models with embedded learning approach, Dian *et al.* [22] proposed the CNN-Fus method, which combines subspace representation and a CNN denoiser for MSI/HSI fusion. The augmented Lagrangian function of the constructed model can be written as:

$$\boldsymbol{\alpha}, V, W = \underset{\boldsymbol{\alpha}, V, W}{\arg\min} \left\| Y - H \boldsymbol{\alpha} \boldsymbol{\psi} \right\|_F^2 + \gamma \left\| Z - \boldsymbol{\alpha} \boldsymbol{\psi} P \right\|_F^2$$
$$+ \rho \left\| V - \boldsymbol{\alpha} + W \right\|_F^2 + \lambda g(V) \qquad (25)$$
$$\text{s.t. } V = \boldsymbol{\alpha}; \quad \text{with } X = \boldsymbol{\alpha} \boldsymbol{\psi}$$

where $\boldsymbol{\psi} \in \mathbb{R}^{L \times S}$ is the subspace according to the subspace estimation [42], which can be estimated from the observed low-resolution MSI. $\boldsymbol{\alpha} \in \mathbb{R}^{WH \times L}$ represents the coefficients. $\boldsymbol{\psi}$ and $\boldsymbol{\alpha}$ are similar to the over-complete dictionary and coefficient in sparse representation, as shown in (4). As in (12), $V$ is the auxiliary variable introduced by the ADMM algorithm.

As shown in Fig. 3, the variables $V$, $\boldsymbol{\alpha}$, and $W$ are iteratively solved. Among these variables, variable $\boldsymbol{\alpha}$ is updated by a fast method based on solving a Sylvester equation [43]. The variable $V$ is updated by the pre-trained CNN denoiser (FFDnet [38] is utilized) as:

$$V = \mathcal{F}\left( \left( \boldsymbol{\alpha}^{t+1} - W^t \right), \frac{\lambda}{2\rho^{t+1}}; \Theta \right) \qquad (26)$$

where $\mathcal{F}$ is the CNN denoiser, $\Theta$ represents the parameters of $\mathcal{F}$, $\boldsymbol{\alpha}^{t+1} - W^t$ is the input of $\mathcal{F}$, and $\frac{\lambda}{2\rho^{t+1}}$ is the noise level $\sigma^2$ in the Gaussian denoiser.

The CNN denoiser is help to supress the noise and the potential artifacts generated in the iteration. Meanwhile, the flexibility of the CNN denoiser guarantee that a well-trained denoising network can be plugged into various image restoration problems, without retraining. However, when the noise of remote sensing images is not significant, denoising prior may not be the optimal option.

To show the effectiveness of the variational models with embedded learning approach in MSI/HSI fusion, in [22], a low-resolution Hyperion MSI with the size of $100 \times 100 \times 89$ and a high-resolution Sentinel-2 MSI with the size of $300 \times 300 \times 4$, with resolutions of 30 m and 10 m, respectively, were adopted in the real-data experiment. The original low-resolution Hyperion HSI has 220 spectral bands in the spectral range of 400–2500 nm, of which 89 bands were retained after removing the bands with a low signal-to-noise ratio (SNR). The original high-resolution Sentinel-2 MSI has 13 spectral bands. Four bands with central wavelengths, i.e., 490 nm, 560 nm, 665 nm, and 842 nm, were utilized in this experiment. Three state-of-the-art fusion approaches were used for comparison with the CNN-Fus coupling method, which includes the nonnegative structured sparse representation (NSSR) [44], coupled spectral unmixing (CSU) [45], and coupled sparse tensor factorization

(CSTF) [46]. Fig. 9 shows the false-color images consisting of the 16th, 5th, and 2nd bands of the fused high-resolution HSIs. As shown in Fig. 9, all the fusion approaches can improve the spatial resolution of the observable low-resolution HSI. The CSU method performs poorly in the spectral fidelity, and obvious artifacts can be seen in the fusion results of NSSR and CSTF. The fusion results of CNN-Fus have much fewer flaws, which shows the superiority of the CNN denoiser compared to the traditional handcraft regularization prior.

### 3) Fusion With the Model-Constrained Network Structure Approach

In the model-constrained network structure approach, Xie *et al.* [47] proposed the MHF-Net method, which unfolds the algorithm into an optimization-inspired deep network for MSI/HSI fusion. The energy function can be written as:

$$\hat{Z} = \underset{\hat{Z}}{\min} \left\| H(ZA + \hat{Z}B) - Y \right\|_F^2 + \lambda f(\hat{Z}) \qquad (27)$$
$$\text{with } X = ZA + \hat{Z}B$$

A novel relational formulation between the observed high-resolution MSI $Z$ and the ideal high-resolution HSI $X$ is introduced as: $X = ZA + \hat{Z}B$, where $\hat{Z} \in \mathbb{R}^{r \times S}$ is the unknown variable to be sought, and $A \in \mathbb{R}^{s \times S}$ and $B \in \mathbb{R}^{r \times S}$ are the corresponding known coefficient matrices. Then the spectral fidelity term between the LR HSI $Y$ and $X$ can be expressed as: $Y = H(ZA + \hat{Z}B)$. More details of this method can be found in [47].

Equation (27) is first decomposed using a proximal gradient algorithm, and then unfolded it into an end-to-end deep network, in which is similar to Fig. 4. Different approaches of this class can adopt different the optimization algorithms. Fig. 4 takes the ADMM algorithm as an example, while in [47] it is the proximal gradient algorithm.

Similar to remote sensing image denoising, the fusion network structure in [47] is designed based on the optimization processing of the objective function to make network interpretable, i.e., (27). Each block of the network represents an iterative solution. Hence, for this approach, the selection of a good network design and a robust optimization strategy can contribute to the admirable performance of this coupling method.

### 4) Fusion With the Model-Constrained Loss Function Approach

In the model-constrained loss function approach, Zhang *et al.* [48] proposed an unsupervised deep framework for blind HSI super-resolution. The loss function in [48] is:

$$\underset{\Theta, P, k}{\min} \left\| Z - X P \right\|^2 + \left\| Y - (X * k)_{\downarrow_r} \right\|^2 + \lambda \left( \left\| k \right\|_2^2 + \left\| P \right\|_F^2 \right) \qquad (28)$$
$$\text{s.t. } X = f_\Theta(E)$$

where $P \in \mathbb{R}^{s \times S}$ denotes the degeneration in the spectral domain. $k$ denotes the unknown blur kernel. $\downarrow_r$ indicates the downsampling operation in the spatial domain with scaling factor $r$, and $*$ indicates the blurring operation in the spatial domain. The first two terms correspond to the spatial fidelity term and the spectral fidelity term. The third term is the regularization term that imposes constraints on the degradation processes: $k$ and $P$, which guides the network to estimate the degradations and improves the ability of the fusion network to comply with the unknown degenerations in real HSI super-resolution applications. $\Theta$-parameterized $f_\Theta(\cdot)$ represents the image generator network for the latent $X$, and $E$ is a precomputed code that contains the image-specific statistics of $X$.

Integrating the observation model into the loss function construction reduces the need for ideal images as label data for the network training, which is very practical and suitable for situations where limited sample data are available.

In general, in remote sensing image fusion, the data-driven and model-driven cascading approach and the variational model with embedded learning approach are simple and easy to implement. The model-constrained network structure approach shows a good performance due to its interpretable structure, but it needs careful mathematical derivation and network structure design. Therefore, there is still not much related work in remote sensing image fusion. Due to the low dependence on ideal high-quality images as label data, the model-constrained loss function approach has been used in some unsupervised fusion methods recently, and it has the potentiality to be combined with the other three combination approaches.

## V. Further Developments

Although the coupled model-driven and data-driven methods have shown their potential in learning the nonlinear correlations in remote sensing imagery and characterizing the structural insight of the optimization based on the image degradation process, challenges still exist due to the complexity of the physical information, the availability of samples, and the universality of the models. In this section, we suggest a few future directions.

### A. Structure Development

#### 1) Introduction of Graph Neural Networks

Two aspects limit the development of the combination strategies in remote sensing applications. Firstly, the convolution operation underpinning all CNN architectures is unable to capture nonlocal self-similarity patterns because of the locality of the convolution kernels. Furthermore, the objects or pixels in multimodal data, including optical data, infrared data, SAR data, LiDAR data, and socioeconomic data, contain complex relationships and interdependencies. The complexity of the data has imposed significant challenges on the existing CNN architectures. Graph neural networks (GCNs) are able to describe the complex relationships between data by the nodes and edges in the feature space of the network [49-51], and have shown great potential for capturing self-similarity information and coping with heterogeneous multimodal data. Very recently, the graph convolutional denoiser network [52] was proposed to aggregate the $k$ nearest neighboring patches for image denoising. In addition, by further mining the multi-scale recurrence property of a natural image, cross-scale internal GCN [53] was proposed to construct the graph on different-resolution patches and successfully recover more detailed textures. However, owing to the big difference between CNNs and GCNs, when the data are structured as a graph, the combination of iterative optimization and GCN is still faced with big challenges. Firstly, the graph-based models are less efficient than the CNN-based models. The graph-based models usually need to handle the whole image, with complex topology structure containing a large number of nodes. For each iteration in the optimization, the graph should updated with changes of an optimal solution. Hence, graph convolutions need to adapt to the dynamicity of the graphs. The change of adjacency relations for each node, however, may also pose a burden to the computational efficiency and introduce turbulence into the network optimization.

#### 2) Utilization of Unsupervised Learning

Most of the existing data-driven image restoration methods are based on supervised learning, i.e., they utilize the ideal clean images as labeled data, and train the network through a large number of samples. However, in actual situations, the following two problems may be encountered. One is the insufficient sample data, such as hyperspectral images. The other is that there may be no ideal clean images as label data of networks. Usually, due to the lack of ideal clean images, scholars have to degrade the observations and use the original observations as ideal clean images. In this case, unsupervised learning that does not require ideal clean images is required. Some scholars have used unsupervised learning for remote sensing image restoration [17, 32, 54, 55]. They have attempted to integrate the observation model in (1) into an unsupervised network, but all of them have mostly adopted the combination form of the model-constrained loss function approach. Thus, the network structure of the unsupervised network still lacks interpretability. In the future, we should consider other model-driven and data-driven combinations in unsupervised learning, especially in the form of the model-constrained network structure approach. Furthermore, in unsupervised learning, it will be valuable to combine the model-driven loss function approach with other approaches.

#### 3) Combination With Tensor Theory

Due to the inherent 3-D characteristics of remote sensing images, the previous vector/matrix-based combination methods have a limited ability to fully exploit the multi-dimensional structural correlation, in comparison with directly working on the high-order tensor format image. The tensor-based methods have received increased attention for preserving the intrinsic structural correlation, obtaining better restoration results, especially the low-rank tensor-based methods over the last two years. The tensor rank is mostly applied to accelerate the deep learning based optimization process [56]. Tensor-based models have been combined with deep learning, but mainly in hyperspectral denoising [31, 57], and they have rarely been used in other remote sensing applications. This combination has been achieved with the variational models with embedded learning, in which a global variational framework is established based on tensor representation, and a deep convolutional network is employed to obtain a plug-and-play prior of ideal images. In the future, on the one hand, tensor models could be embedded into deep learning using other combination methods, providing more physical constraints for the data-driven methods. On the other hand, tensor decomposition could be used to obtain more essential feature components of the data, which could be extracted or updated by network training.

### B. Application Challenges
#### 1) The Complex Degradation Problem

Thin cloud, haze, and shadow often cause unevenness in remote sensing images, which can influence human interpretation. Haze/thin cloud degradation is triggered by the scattering of radiance by the turbid particles in the atmosphere, which makes the ground information distorted. Shadow leads to an intensity decrease due to the obstruction of the incident light. Therefore, it is essential to correct the unevenness. Many deep learning based cloud/shadow removal methods have been proposed, but only a few are combined with variational models [14] [58] to handle natural images. For example, Yang *et al.* [14] unfold the iterative algorithm with the transmission prior and dark channel prior to be a deep network; and Liu *et al.* [58] formulated image dehazing as the minimization of a variational model with favorable data fidelity terms and prior terms, and then solved the variational model based on the classical gradient descent method with built-in deep CNNs. However, no model-driven deep learning methods have been proposed to remove thin cloud, haze, and shadow from remote sensing images. Compared to natural images that are covered with uniform haze or shadow, the actual intensity in remote sensing images is usually non-uniform, and is thus more complex in the spatial and spectral domains. Two possible alternative ideas are put forward here to solve the problem. One possibility is to simulate the non-uniform and complex degenerations through a convolutional layer rather than simple image statistics. Secondly, although the unevenness caused by haze, thin cloud, or shadow has no obvious distribution law in the spatial domain, a strict physical scattering law related with the intensity of spectral bands can be incorporated into deep learning to improve the solution [59-61].

#### 2) Large Area Missing Information

Due to sensor malfunction and adverse atmospheric conditions, there can often be a great deal of missing information in optical remote sensing data. This makes missing information reconstruction technology important. Recently, some researchers have devoted efforts to develop the inpainting based on the coupled model-driven and data-driven methods [34, 62]. For example, Sidorov *et al.* [34] used the intrinsic properties of a CNN without any training to obtain the regularization terms in a variational model; and Lahiri *et al.* [62] first trained a generative model to map a latent prior distribution to the natural image manifold and search for the 'best-matching' prior to reconstruct the signal. However, in remote sensing images, the land-cover types are complex, especially for high-resolution images. Moreover, thick cloud is usually accompanied with cloud shadows, which can be further divided into umbra and penumbra. This makes the inpainting for remote sensing images face serious challenge.

#### 3) Heterogeneous Image Fusion

The existing model-driven and data-driven combinations are mostly adopted for single-type image restoration, such as SAR image denoising or optical image fusion. The fusion of heterogeneous images with different statistical properties can, in theory, improve the performance of remote sensing images, which is meaningful. In fact, there have already been some studies of heterogeneous image fusion [63-67]. Most of the early heterogeneous image fusion methods were based on a simple linear assumption [64,65]. Since then, some heterogeneous fusion methods based on the model-driven approach [66,67] have gradually been developed. However, due to the difference in the imaging mechanisms between heterogeneous images, it is difficult to construct an accurate model. In recent years, with the rapid development of deep learning, some scholars have turned to the use of deep learning for heterogeneous image fusion [68-70]. In heterogeneous image fusion, the coupling of model-driven and data-driven methods has not yet been attempted, and there is much research space.

## VI. CONCLUSIONS

In this paper, we have systematically reviewed the combinations of model-driven and data-driven methods for remote sensing image restoration and fusion. The combination methods solve the black-box problem of deep learning methods by introducing the model-driven method to direct the parameter learning and improve the physical interpretability of the model. The coupling approaches can be further divided into three types: 1) data-driven and model-driven cascading; 2) the variational models with embedded learning; and 3) model-constrained network learning. These techniques have been widely applied in remote sensing image restoration and fusion, especially in SAR image despeckling, HSI hybrid noise reduction, and remote sensing image fusion. The results described in this paper confirm that the use of these approaches can result in a significant improvement for remote sensing image restoration and fusion.

However, research into the coupling of model-driven and data-driven methods is still young. We believe that some new insights into the potential improvement for remote sensing applications have been provided in this paper. From the perspective of the model structure, GCN and deep tensor decomposition [71] can be integrated into a combination framework for better utilization of the spatial nonlocal self-similarity patterns and high spectral correlation property. To remove the dependency on large numbers of training samples, how to achieve transfer learning between datasets and unsupervised learning without labeled HR images will be essential tasks in the future. On the other hand, the more diverse application directions are worth studying, such as dehazing, cloud removal, heterogeneous image fusion. Exploring proper data-driven priors based on an optimization-inspired variational mode for these more complex and specific remote sensing problems also remains a big challenge.